\documentstyle[prl,aps,multicol,epsf]{revtex}
\input psfig

\newcommand{\ve}{\varepsilon}
\newcommand{\br}{{\bf r}}

\newcommand{\dU}{\delta U}

\begin{document}
\setlength{\unitlength}{3in}
\title{Acoustoelectric current and pumping
in a ballistic quantum point contact}
\author{Y. Levinson }
\address{ Department of Condensed Matter Physics, The Weizmann 
Institute of Science, Rehovot 76100, Israel}
\author{O. Entin-Wohlman}
\address {School of Physics and Astronomy,
Raymond and Beverly Sackler Faculty of Exact Sciences,\\
 Tel Aviv University, Tel Aviv 69978, Israel}
\author { P. W\"olfle}
\address{Institut f\"ur Theorie der Kondensierten Materie,
Universit\"at Karlsruhe, 76128 Karlsruhe, Germany}
\date{\today}
\maketitle
\begin {abstract}
The acoustoelectric current 
induced by a surface acoustic wave (SAW) in a ballistic
quantum point contact is considered using
a quantum approach.  We find that
the current is of the ``pumping" type and is not related to
drag, i.e. to the momentum transfer from the wave to the electron gas.
At gate voltages corresponding to the plateaus of the quantized
conductance the current is small. It is peaked at the conductance step 
voltages.
The peak current oscillates and decays with increasing SAW
wavenumber 
for short wavelengths. These results contradict previous calculations, 
based on the classical Boltzmann equation.
\end {abstract}
\pacs {PACS numbers: }
\begin{multicols}{2}

The interaction of surface acoustic waves (SAW) with electrons
in a two dimensional electron gas
(2DEG) has recently attracted much attention.
In particular the acoustoelectric effect 
(d.c. current driven by the SAW)
was investigated experimentally in a point contact (PC) 
defined in GaAs/AlGaAs heterostructure by a
split gate \cite{Sh1,Sh2,Ta97}.
Most of the 
 theoretical considerations of this effect were classical,
based on the Boltzmann equation for electrons in a 1D channel,
with the SAW considered as a classical force \cite{Sh1,To96}
or as a flux of monochromatic surface phonons\cite{Gu96,Gu98}.
Such an approach is valid only when the channel length
is much longer than the electron Fermi wavelength and when the electron
diffraction
at the channel ends can be neglected.
In this picture the acoustoelectric current results
from the drag of electrons by the SAW. Its value
is determined by the competition between the momentum transfer
from the SAW to the 2DEG and the momentum relaxation 
due to impurity scattering \cite{Sh1,To96} 
or due to electron escape from the PC \cite{To96,Gu96,Gu98},
for a ballistic PC. 

A quantum approach  was used in \cite{Ma97},
but only PC's of length
short compared to the SAW wavelength
were  considered for the experimentally relevant low frequencies.
We present here a quantum description of the problem,
based on a different formalism, which allows a
more general consideration and leads
to results qualitatively different from those given by 
the classical approach. In particular, we find that the drag mechanism is not
valid for the quantum acoustoelectric current, and that the reflection of
the elctrons within the PC is crucial for producing the SAW effect.
Unfortunately the results of the experiments do not allow to 
reach a definite conclusion about the mechanism of the 
acoustoelectric current.

Consider a nanostructure (NS) of arbitrary geometry 
(e.g. a PC) where the 2DEG is confined by a
potential $U(\br)$ and is attached to terminals $\alpha$
(with no voltage bias). 
The NS is exposed to a random a.c. potential $\delta U(\br,t)$,
produced by a gate or by radiation, infrared or acoustic, and 
localized within the NS. This a.c. potential induces a current
through  the NS, the d.c. component of which is the acoustoelectric current
or the photovoltaic current, depending on the nature of the radiation.
Using {\it time-dependent} scattering states 
(see below for details), we find that
the d.c. current $J_{\beta}$ entering terminal $\beta$
is given by $(e<0,\hbar =1)$
\begin{eqnarray}
\label{J}
J_{\beta}=\sum_{\alpha}J_{\beta\alpha},\qquad
J_{\beta\alpha}=-e
\int\int d\br _{1} d\br _{2}P(\br_{1},\br_{2})\times
\nonumber\\
\int{ dE\over2\pi}\left(-{\partial f(E)\over \partial E}\right)
 g_{\beta}(E|\;\br_{1},\br_{2})
g_{\alpha}(E|\;\br_{1},\br_{2}).
\end{eqnarray}
The properties of the a.c. potential are  condensed in
the pumping factor
\begin{eqnarray}
\label{P}
P(\br_{1},\br_{2})=
\int d\omega\omega\overline{\dU (\br _{1}) \dU (\br _{2})}^{\omega},
\end{eqnarray}
where $\overline{\dU (\br _{1}) \dU (\br _{2})}^{\omega}$
is the Fourier component of the random field correlator
$\overline{\dU (\br _{1},t_{1}) \dU (\br _{2},t_{2})}$
and the  overbar denotes statistical averaging.
The properties of the NS are embodied in
\begin{equation}
\label{g}
g_{\alpha}(E|\br_{1},\br_{2})=
\sum_{n}\chi_{\alpha n}(E|\br_{1})\chi_{\alpha n}(E|\br_{2})^*,
\end{equation}
where $\chi_{\alpha n}(E|\br)$ is a scattering state
excited by an incoming wave $w^{-}_{\alpha n}(E|\br)$
 (normalized to unit incoming flux) of an
 electron with energy $E$ entering the NS from channel 
$n$ of terminal $\alpha$. 
$f(E)$ is the Fermi distribution in the terminals.

Equation (\ref{J}) is valid in the weak field  adiabatic approximation,
when the a.c. perturbation $\delta U$ is smaller than the Fermi 
energy $E_{F}$ and
when the relevant frequencies of this perturbation  are
smaller than all scales which determine the energy dependence of the
scattering states. 
The statistical averaging replaces the temporal averaging,
which is unavoidable when measuring a d.c. current
induced by an a.c. perturbation. 
Below we assume zero temperature,
which reduces the energy integration in (\ref{J}) to 
$E=E_{F}$.

The a.c. potential created by a SAW 
propagating in the $x$ direction is 
$\delta U(\br,t)=A(t)\exp i(qx-\omega_{0}t)+c.c.$, where
the amplitude $A(t)$ is a stationary, slowly varying, random function.
This potential is screened by the electrons of the 2DEG
\cite{Kn97}.
In the wide part of the PC  
 the screening strongly reduces the potential (by 
the factor $qa_{B}$, where $a_{B}$ is the Bohr radius),
while in the narrow part 
screening is weak.
To account for the screening effect we  multiply
$\delta U(\br,t)$ by a screening  factor $S(x)=S(-x)$.
 For this screened a.c. potential  
the pumping factor is
$P(\br_{1},\br_{2})=2i\omega_{0}\overline{|A|^{2}} \sin q(x_{1}-x_{2})
S(x_{1})S(x_{2})$ . 

Let us first consider a PC attached to two ideal 1D leads at 
$x\rightarrow \pm\infty $, and assume that the PC is represented by a repulsive
delta function potential, $U(x)=V\delta(x)$.
The scattering states excited from the left terminal 
$\alpha=l$ at $x=-\infty$ and the right terminal $\alpha=r$ at $x=+\infty$
are  
$\chi_{\alpha}(E|x)= 
v_{E}^{-1/2}[\exp(\pm ik_{E}x)+r_{E}\exp(ik_{E}|x|)]$, where $\pm $
denotes $\alpha=l$ and 
$\alpha=r$, respectively, and  $k_{E}$ and $v_{E}$ are the
electron momentum and velocity at energy $E$.
The transmission and reflection amplitudes are
$t_{E}=1+r_{E}=(1+iV/v_{E})^{-1}$. 
For the screening factor we choose $S(x)=\exp(-|x|/L_{s})$.
Eq. (\ref{J}) then yields that the partial
currents $J_{lr}=J_{rl}=0$, while $J_{ll}=-J_{rr}\equiv J$,
where $J$ is the d.c. current through the PC in the $x$-direction.
For $q\ll k_{F}$ we have
\begin{equation}
\label{J1D}
J=e{\omega_{0}\over 2\pi}\;{\overline{|A|^{2}}\over 2E_{F}^2}\;
{qk_{F}\over q^2+L_{s}^{-2}}\;|t_{F}|^2|r_{F}|^2,
\end{equation}
where the index $F$ means $E=E_{F}$. 
This result shows that (i) the current is proportional to
$\omega_{0}$, and hence is of the pumping type; (ii) the current increases with
$q$ for small wavevectors, and decays for large ones; (iii) a finite reflection
is necessary for producing the effect.

Turning now to a more realistic description of the PC, we model it by
the 2D saddle-point potential
$U(x,y)=(1/2md^{2})[-(x/L)^2+(y/d)^2]$,
where $m$ is the electron mass, 
$L$ is the length of the PC and $d$ is its width.
For $L\gg d$ this potential corresponds to a waveguide in the $x$-direction
with parabolic walls (at $|x|\alt L$) adjusted to horns
(at $|x|\agt L$) with opening angle $d/L$.
These horns represent the left and right terminals at
$x=\mp \infty$. 
The scattering states
are given by \cite{Le92}
$\chi_{\alpha n}(E|\br)=\Phi_{n}(y)\chi^{\pm}(\ve_{n}|x)$.
Here $\Phi_{n}$ is a normalized harmonic oscillator wave function
with energy $E_{n}=\Delta(n+1/2)$, where  
$\Delta=1/md^2$, and $n=0,1,2,...$ labels the channels
in both terminals. (There is no channel mixing in a 
saddle-point potential.) $\chi^{\pm}(\ve_{n}|x)$
is given by the complex Weber
(parabolic cylinder) function
$E(a,x)$, as defined in \cite{Ab64}
\begin{equation}
\chi^{\pm}(\ve_{n}|x)=
-i \sqrt{m}(Ld/2)^{1/4}t(\ve_{n})E(-\ve_{n},\pm\xi).\label{CHIPM}
\end{equation}   
Here $\xi=(2/Ld)^{1/2}x$ and 
$\ve_{n}=(E-E_{n})/\delta$ with $\delta=1/mLd$. In Eq.(\ref{CHIPM})
$t(\ve)=(1+e^{-2\pi\ve})^{-1/2}$ is the transmission amplitude of the
barrier created by the saddle-point potential. 
Again, $\pm $ denote $\alpha=r,l$, respectively.

The Landauer conductance (at zero temperature, in units of $e^2/h$) of such a PC
is  \cite{Bu90} ${\cal G}=\sum_{n}t(\ve_{n})^2$,
where now $\ve_{n}=(E_{F}-E_{n})/\delta$. When
$L\gg d$ the conductance as a function of $E_{F}$ is a step like function, with
plateaus of width $\Delta $ and steps of width $\delta $.
The steps occur at energies $E_{n}$ where $E_{F}$ equals the bottom
of the transverse quantization mode $n$; For
$E>E_{n}$ this mode is propagating,
while at  $E<E_{n}$ it is evanescent.

The current in the saddle-point PC, as obtained from Eq. (\ref{J}),
consists of a sum over the
separate mode contributions
\begin{eqnarray}
\label{Jac}
J=J_{0}\sum_{n}F(\ve_{n},p),\ \ p=q(Ld/2)^{1/2},
\end{eqnarray}
with the nominal value 
$J_{0}=2e(\omega_{0}/2\pi) (\overline{|A|^{2}}/\delta ^2)$.
The function $F(\ve,p)$ [see Eq. (\ref{FGH}) below] 
is positive for $p>0$, and is odd in
$p$, i.e. the electron flux is along the
direction of the SAW propagation. We find that
$F(\ve,p)$ is exponentially small when $|\ve|\gg 1$, that is, modes whose
energies are far from the threshold $E_{n}$
do not contribute to the current.
This is expected for the evanescent modes;  for the propagating ones
it means that in a free channel with no reflection
the acoustoelectric current is zero.
The crucial role of reflection in producing the current can be seen
also from Eq. (\ref{J1D}).

Near the threshold, for $|\ve|\alt 1$,
where the current is not small, we have (for $p>0$)
\begin{eqnarray}
\label{F}
F(\ve,p)&=&2\pi e^{-\pi\ve}t(\ve)^3 c(\ve)
{\rm erf}\left({p\over \sqrt{2\sigma}}\right),\qquad p\ll 1,\nonumber\\
F(\ve,p)&=
&4\pi t(\ve)^2 \cos^2\left({p^2\over 2}-{\pi\over 4}-\gamma_{\ve}\right)
{e^{-\sigma p^2}\over  p^2},\ p\gg 1.
\end{eqnarray}
Here  $\sigma=Ld/L_{s}^2,\;
\gamma_{\ve}=2\ve\ln p+\arg\Gamma(1/2-i\ve)$
and $c(\ve)\simeq 1$ is given by integral $H$  [see Eq. (\ref{GH}) below] 
at $p=0,\;\sigma=0$.
In this calculation the screening factor is chosen to be  
$S(x)=\exp(-x^2/L_{s}^2)$. Typically
$L_{s}=L$, which results in
$\sigma=d/L\ll 1$. 

It follows from Eqs. (\ref{Jac}) and (\ref{F})
that for long SAW waves  $q\ll L^{-1}$
 the current increases linearly with $q$.
For $L^{-1}\ll q\ll (Ld)^{-1/2}$ the current is independent of $q$
and for short waves $(Ld)^{-1/2}\ll q$ it exhibits damped oscillations.
It is interesting to note that the oscillations are not simple geometrical; 
the wavelength ``resonates'' not with the channel length $L$,
but with the less obvious length  $(Ld)^{1/2}$. The numerical calculations 
of a single mode contribution to the  current are depicted in the figure,
for $L/d=10$ and $\ve=0,\pm $0.5. The intermediate region in which $F$ is
independent of $q$ is not distinguished for the
not very small $d/L$ ratios. One can see from this figure that
below the threshold ($\ve<0$)
the current is much weaker then above it ($\ve>0$).
\vspace{-2cm}
\narrowtext
    \begin{picture}(1,1)
\put(0,0){\psfig{figure=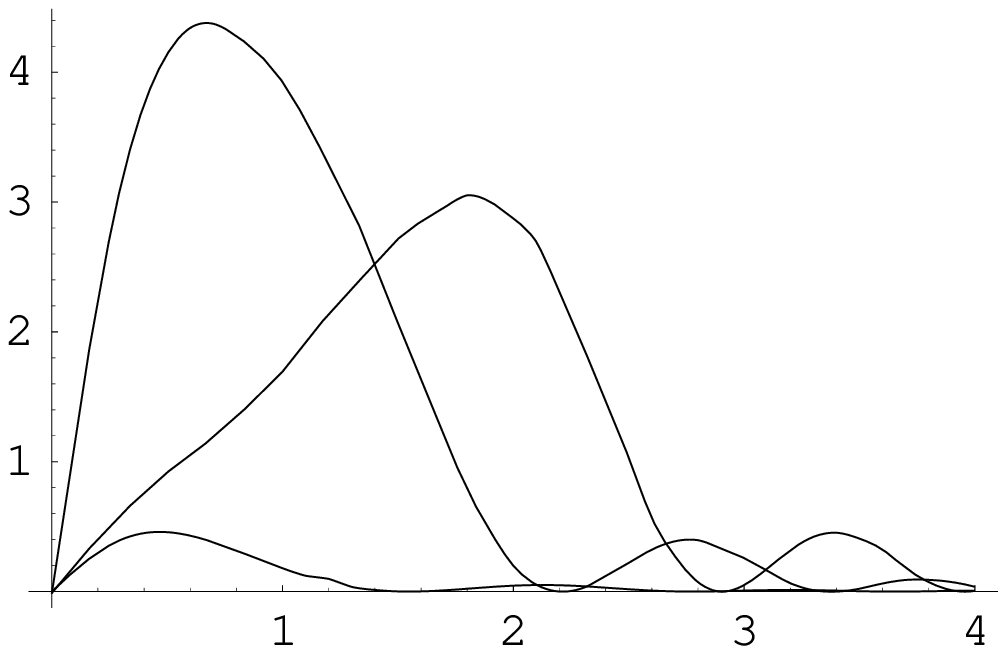,width=3in}}
\put(0,-.07){Fig.1 Contribution of a single channel to the current}

\put(0,-.13){(see Eq. (6)) as a function of the dimensionless SAW }
\put(0,-.19) {wavenumber $p$, for different values of Fermi energy} 
\put(0,-.25){close to the step of the conductance}
\put(1.05,.05){\makebox(0,1)[lb]{$p$}}
\put(0,.67){\makebox(0,1)[lb]{$F(\varepsilon,p)$}}
\put(.31,.55){\makebox(0,1)[lb]{$\varepsilon=0$}}
\put(.6,.3){\makebox(0,1)[lb]{$\varepsilon=+0.5$}}
\put(.19,.14){\makebox(0,1)[lb]{$\varepsilon=-0.5$}}
    \end{picture}

\vspace{2cm}

Our quantum theory predicts that the current is strong only when
$E_{F}$ is at the vicinity
of the transverse quantization channel bottom. 
This finding is in agreement with
that of the classical approach \cite{Sh1,To96,Gu96,Gu98}, of giant
oscillations in the acoustoelectric current.
However, in the quantum theory the width of the peak, $\delta $,  
is determined by the
diffraction at the opening  angle of the channel. Indeed, the width-to-spacing
ratio of the oscillations is $\delta /\Delta =d/L$. On the other hand, the
peak width in the classical theory is determined (at zero temperature)
by the scattering or by the escape time in the case where 
the SAW is described as a 
force, or by the energy and momentum conservation when the SAW
is described as a phonon flux.
Experiment \cite{Sh1} shows the oscillations, however
unfortunately the peaks are not very pronounced, and this is why
one cannot make statements about the nature of their width.

 Another important difference between the quantum
and the classical approaches concerns the behavior of the current 
at short waves.
According to the (classical) drag picture, the current should increase 
with $q$,
whereas the quantum theory yields exponential decreasing.
(Note that in these considerations the proportionality of
$\omega_{0}$  to $q$ is not relevant.)
For a short PC our theory predicts that $J/J_{0}\simeq qL$.
This linear dependence on $q$ was also  obtained in Ref. \cite{Ma97}
[see Eq. (19) there], however the coefficient was not specified.
 Estimating 
in that equation $\partial _{\mu}T^{0}(\mu)\simeq 1/\delta$ 
(where  $T^{0}(\mu)$ is the transmission at $E_{F}$ ) 
the current (in our notation) becomes
$J/J_{0}\simeq qL\times g (L/v_{{\rm SAW}})\delta$,
where $g$ is a geometry factor
independent on $L$ \cite{Ma97}.
 Since the ratio $(L/v_{{\rm SAW}})\delta$
is also $L$-independent, the predictions of both theories
agree regarding the dependences on SAW frequency and contact length.

We now outline the derivation of Eq. (\ref{J}). This is accomplished using
the concept of {\it time-dependent}
scattering states \cite{Le99}. Let the NS under the a.c. field
be described by
the Hamiltonian
${\cal H}=\int d\br \Psi^{+} (\br)H(\br,t)\Psi (\br)$, where
$H(\br,t)=H_{0}(\br)+\dU (\br,t)$ and  
$H_{0}(\br)=
(1/ 2m)(-i\nabla)^2+U(\br)$.
Here $\Psi (\br)$ is the electron field operator.  
The time-dependent scattering state $\chi_{\alpha n}(E|\br,t)$ 
is  defined as the solution of the equation
\begin{equation}
i(\partial/\partial t)\chi_{\alpha n}(E|\br,t)
=H(\br,t)\chi_{\alpha n}(E|\br,t),
\label{HCHI}
\end{equation}
 which is  excited by
an incoming wave $w^{-}_{\alpha n}(E|\br)\exp(-iEt)$
{\it in the presence of the a.c. potential}.
The state $\chi_{\alpha n}(E|\br,t)$ is labeled according to the energy
of the  incoming wave, but it contains components
with energies $E'\neq E$,
since due to the time dependent perturbation $\dU (\br,t)$ the
transmission and the reflection of the incoming wave are inelastic.
For a weak time-dependent potential, $\delta U\ll E_{F}$,
this equation can be solved by iterations,
\begin{eqnarray}
\label{chiexp}
\chi_{\alpha n}(E|\br,t)=
\qquad \qquad \qquad \qquad\qquad \qquad \qquad 
\\ \nonumber
 e^{-iE t}
[\chi_{\alpha n}(E|\br)
+\chi_{\alpha n} ^{(1)}(E|\br,t)+
\chi_{\alpha n} ^{(2)}(E|\br,t)+...]. 
\end{eqnarray}
The first term here is the (time-independent) scattering solution of 
$H_{0}$, and
the subsequent terms contain only outgoing waves. The latter can be found 
in terms of the retarded  Green's function of $H_{0}$,
\begin{eqnarray}
\label{Gf}                  
(i\partial/\partial t -H_{0}(\br))
G(\br,\br ',t)=\delta(\br-\br ')\delta(t).
\label{dGxt}
\end{eqnarray}

The time-dependent field operator, required for the calculation of the current
density operator
${\bf j}(\br,t)=-(ie/2m) \Psi(\br,t)^{+} \nabla \Psi(\br,t) + h.c.$, 
can be written
in terms of the scattering states,
\begin{eqnarray}
\Psi (\br,t)=\int{dE \over 2\pi}
\sum_{\alpha n}a_{\alpha n}(E)\chi_{\alpha n}(E|\br,t).
\label{fo}
\end{eqnarray}
Here $a_{\alpha n }^{+}(E)$ is an operator creating an
incoming electron in channel $n$ of lead  $\alpha$
with energy $E$.
The averages of these operators are determined by the temperatures
and the chemical potentials of the terminals connected to the leads.
For the scattering states which are normalized to unit
incoming flux
\begin{eqnarray}
\label{aa}
\langle a^{+}_{\alpha n }(E)  a_{\alpha ' n'}(E') \rangle
=2\pi\delta(E -E')\delta_{\alpha n,\alpha ' n'}f_{\alpha}(E),
\end{eqnarray}
where $f_{\alpha}(E)$ is the Fermi distribution
in terminal $\alpha$.

Using the above results one performs the quantum and statistical averaging 
to obtain the current density $\overline {\langle{\bf j}(\br)\rangle}$.
Evaluating $\overline {\langle{\bf j}(\br)\rangle}$ far away 
in terminal $\beta$ and integrating 
over the cross section of this terminal gives $J_{\beta}$ of Eq. (\ref{J}).
The asymptotic behavior of the current density is derived using the
following relation \cite{Le00} 
for the Fourier transform of the Green function
defined by Eq.(\ref{Gf}),
\begin{equation}
\label{Ginf}
G(E|\br,\br')|_{\br\rightarrow \infty\beta}=
-i\sum_{m}w_{\beta m}^{+}(E|\br) \chi_{\beta m }(E|\br'),
\end{equation}
where $w_{\beta m}^{+}(E|\br)$ is an outgoing wave in channel
$m$ of terminal $\beta$ (normalized to unit flux).

The acoustoelectric current for the saddle-point confining potential, 
Eq. (\ref{Jac}), is obtained using for the complex Weber functions the
representation \cite{Ab64}
$E(a,x)=k^{-1/2}W(a,x)+ik^{1/2}W(a,-x)$, where $W(a,\pm x)$
are the real Weber functions and $k=(1+e^{2\pi a})^{1/2}-e^{\pi a}$.
We find
\begin{eqnarray}
\label{FGH}
F(\ve,p)=t(\ve)^{3}G(\ve,p)H(\ve,p),
\end{eqnarray}
where
\begin{eqnarray}
\label{GH}
G(\ve,p)=\int_{-\infty}^{+\infty}d\xi \exp(-\sigma\xi^2/2)
\sin p\xi\; g(\ve,\xi),\nonumber\\
H(\ve,p)=\int_{-\infty}^{+\infty}d\xi \exp(-\sigma\xi^2/2)
\cos p\xi\; h(\ve,\xi),
\end{eqnarray}
with
\begin{eqnarray}
\label{gh}
g(\ve,\xi)=W(-\ve,-\xi)^2-W(-\ve,\xi)^2=-g(\ve,-\xi),\nonumber\\
h(\ve,\xi)=W(-\ve,\xi)W(-\ve,-\xi)=h(\ve,-\xi).
\end{eqnarray}
The appearance of the transmission amplitude $t(\ve)$ in Eq. (\ref{FGH}) 
ensures the exponential smallness of the
function $F(\ve,p)$ for evanescent modes. 
To show that it is also exponentially  small for
propagating modes belonging to $\ve\gg 1$ we use the 
Darwin representation of the Weber
functions \cite{Ab64} $ (\xi>0)$ and obtain
\begin{eqnarray}
\label{D}
g(\ve,\xi)=\left[\sqrt{\ve}\cosh(s/2)\right]^{-1}
[e^{-\pi\ve}+\sin\theta],
 \nonumber\\
h(\ve,\xi)=\left[2\sqrt{\ve}\cosh(s/2)\right]^{-1}\cos\theta,
 \nonumber\\
\theta=\ve(s+\sinh s),\; \xi=2\sqrt{\ve}\sinh (s/2).
\end{eqnarray}
The functions $g$ and $h$ then
contain exponentially small or  fast oscillating terms. 
The  integrals $G$ and $H$ can then be calculated near the saddle-point
$s=i\pi$ and are found to be $\sim e^{-\pi\ve}$.

We now turn to the case $|\ve|\alt 1$. 
For large $p$, $F(\ve,p)$ is determined by the singular points of  
$g$ and $h$. These functions
are regular  on the real $\xi$ axis;
at $\xi\rightarrow\infty$ they are given by
\begin{eqnarray}
\label{ghinf}
g(\ve,\xi)=2\xi ^{-1}[e^{-\pi\ve}+t(\ve)^{-1}\sin\vartheta],
 \nonumber\\
h(\ve,\xi)=\xi^{-1}\cos\vartheta,
\nonumber\\
\label{phi}
\vartheta=\xi^2/2+2\ve\ln\xi+\arg\Gamma(1/2-i\ve).
\end{eqnarray}
Thus, the main contribution to 
$H$  and $G$  comes from large $\xi$,
as both $g$ and $h$ have a singular point
$\xi=\infty$. Using Eq. (\ref{ghinf}) one can check that
the saddle points of $\cos\vartheta \cos p\xi$
and $\sin\vartheta \sin p\xi$ are  $\xi=\pm p$. Calculating
$G$ and $H$ near these points yields the second of Eqs. (\ref{F}).

For small $p$ the behavior of 
$G$ and $H$ is different. In $H$ one can put 
$\cos p\xi=1$ and $\sigma=0$. In $G$, however, the limit 
$p\rightarrow 0$ should be taken with care: for $\sigma=0$, 
$G$ has a singularity of the form ${\rm sgn}(p)$ coming from the 
non-oscillating term in
$g$. The factor erf$(p/\sqrt{2\sigma})$ 
in the first of 
Eqs. (\ref{F}) results from the smoothing of this singularity by the screening
factor.

Being proportional to the frequency $\omega_{0}$ of the SAW,
the acoustoelectric 
current is of the pumping type: independent of the value of $\omega_{0}$,
a given fraction of the electron charge is transferred through
the PC  during each period of the SAW.
Therefore, the current can be compared with the pumping
current produced by two gates with phase shifted a.c. potentials,
$\delta U(\br,t)=A_{1}(t)u_{1}(\br)+A_{2}(t)u_{2}(\br)$.
Let the gates be symmetric, 
$u_{1}(\br)=u(x),u_{2}(\br)=u(-x)$ and take
$A_{1}(t)=A(t)\cos(\omega_{0}t+\varphi(t)+\phi),
\;A_{2}(t)=A(t)\cos(\omega_{0}t+\varphi(t)) $,
where the amplitude $A(t)$ and the phase $\varphi(t)$ 
are slowly varying, random functions,
but the phase shift $\phi$ is fixed.

One can compare now the pumping factors, Eq. (\ref{P}),
for the SAW and the two gates and see that they are
equal if one replaces
$\exp(iqx)$ by $ [\exp(i\pi/4)u(x)+\exp(-i\pi/4)u(-x)]/\sqrt{2}$.
It  means that a propagating SAW  is equivalent
to  pumping
with a phase shift $\phi=\pi/2$ between two symmetric gates.
 This is why the acoustoelectric current, although
being of pumping type, does not contain the factor $\sin\phi$,
typical for pumping \cite{He91,Br98}.
  
This work was supported by the Alexander von Humboldt Foundation,
the Israel Academy of Sciences
(Y.L), the German-Israeli Foundation and the Deutsche 
Forschungsgemeinschaft (P.W).

\end{multicols}
\end{document}